%% file: main.tex
\documentclass%
[%
10pt,
oneside
]{article}

\usepackage[a4paper, margin=1in]{geometry}
\usepackage[ruled,vlined]{algorithm2e}

\include{preamble}

\normalfont

\begin{document}

\begin{titlepage}
    \begingroup
    \vspace*{1cm}
    \Large
    \noindent
    \textbf{Directed Acoustic Assembly in 3D}
    
    \normalsize
    \raggedright
    Kai Melde\ts{1,2}, Minghui Shi\ts{1,2}, Heiner Kremer\ts{3}, Senne Seneca\ts{2,4}, Christoph Frey\ts{2,4}, Ilia Platzman\ts{2,4},\\
    Christian Degel\ts{5}, Daniel Schmitt\ts{5}, Bernhard Sch\"olkopf\ts{3} and Peer Fischer\ts{1,2*}\\
    \vspace{0.5cm}

    % AFFILIATIONS
    %\smaller
    \ts{1} Micro, Nano and Molecular Systems Group, Max Planck Institute for Medical Research, Jahnstr. 29, 69120 Heidelberg, Germany\\
    \ts{2} Institute for Molecular Systems Engineering and Advanced Materials, Heidelberg University, Im Neuenheimer Feld 225, 69120 Heidelberg, Germany\\
    \ts{3} Empirical Inference Department, Max Planck Institute for Intelligent Systems, Max-Planck-Ring 4, 72076 T\"ubingen, Germany\\
    \ts{4} Department of Cellular Biophysics, Max Planck Institute for Medical Research, Jahnstr. 29, 69120 Heidelberg, Germany\\
    \ts{5} Technical Ultrasound Department, Fraunhofer Institute for Biomedical Engineering, Ensheimer Straße 48, 66386 St. Ingbert, Germany\\
    \vspace{0.1cm}
    \ts{*} corresponding author: peer.fischer@mr.mpg.de\\
    \vspace{0.5cm}
    \endgroup
    
    \normalsize
    \section*{Abstract}
    The creation of whole 3D objects in one shot is an ultimate goal for rapid prototyping, most notably biofabrication, where conventional methods are typically slow and apply mechanical or chemical stress on biological cells. Here, we demonstrate one-step assembly of matter to form compact 3D shapes using acoustic forces, which is enabled by the superposition of multiple holographic fields. The technique is contactless and shown to work with solid microparticles, hydrogel beads and biological cells inside standard labware. The structures can be fixed via gelation of the surrounding medium. In contrast to previous work, this approach handles matter with positive acoustic contrast and does not require opposing waves, supporting surfaces or scaffolds. We envision promising applications in tissue engineering and additive manufacturing.

    \vspace{0.5cm}
    \noindent
    \textbf{Keywords:} one-step, directed assembly, acoustic hologram, radiation force\\
    
    \vfill

\end{titlepage}

\section*{Introduction}
One-shot fabrication of polymer structures has recently been proposed as a manufacturing concept, in which a whole three-di\-men\-sion\-al object is formed in one shot~\cite{RN426}. In this case the object's 3D shape has to be defined by a suitable intensity distribution of the generating light field, e.g. by a photo-initiated polymerization reaction. However, it is difficult to shape optical fields in a compact 3D volume at once using conventionally available devices and spatial light modulators. We refer to \emph{compact} 3D fields as those, which exhibit feature sizes over similar length scales in all three dimensions. For a projected image the achievable lateral resolution scales with \(1/\mathrm{NA}\) and the axial resolution with \(1/\mathrm{NA}^2\), where \(\mathrm{NA}\) is the numerical aperture~\cite{RN619}. Thus, to create compact 3D holographic images the optimal working distance should be about half the aperture diameter, so that \(\mathrm{NA}\approx 1\). Such high numerical apertures are thus far only practicable for illuminating microscopic target volumes. Examples for optical computer generated holography (CGH) at larger dimensions have correspondingly lower NAs and the resulting images are stretched by several orders of magnitude in the beam propagation direction~\cite{RN586,RN608}, which limits their use for fabrication purposes of compact objects. Proposed solutions to this problem are thus far based on the nonlinear activation of a photoinitiator. One promising approach is the sequential illumination from varying angles into a rotating chamber to obtain projected tomography~\cite{RN570,RN643,RN571}. In this case it is necessary that the chemical crosslinker activates only when the accumulated light dose exceeds a certain threshold~\cite{RN571}. Recently, this concept has been demonstrated to also work within lightly scattering media such as cell suspensions and using a cytocompatible photoinitiator~\cite{RN717}. An alternative approach, called Xolography~\cite{RN612}, is based on two-photon excitation with different wavelengths and consequently allows separation into two beams: one illuminating a plane with a focused lightsheet and the second beam projecting the corresponding cross section of the object on that plane. Since both excitation wavelengths are necessary, the polymerization reaction is only initiated in one plane at a time. It stands to reason that both methods, however, still generate the 3D fields in a serial manner and not in a single step. 

An alternate approach to light is to use sound fields that exert acoustic forces for assembly. This approach has the distinct advantage that the material of interest, such as particles or cells, is directly manipulated. Furthermore, the application of ultrasound at the intensities used here is cytocompatible and does not warrant the use of chemical additives such as photoinitiators.  Indeed, acoustic particle assembly shows promise for rapid prototyping~\cite{RN424} and application in cell cultures~\cite{RN610}, but has thus far been limited to 2D assembly close to boundaries~\cite{RN300,RN727} or point-like tweezing in air~\cite{RN244,RN515} and water environments~\cite{RN255,RN408}. These tweezing methods require phase discontinuities in the focal region, which complicates their extension towards extended traps. Alternatively, standing waves can be used to assemble cells in 2D into regular patterns~\cite{RN484,RN735}, or colloidal microparticles in 3D~\cite{RN220}, but here the patterns are generally highly symmetric. Prisbrey et al. have examined the use of a closed cavity whose surface is fully lined by transducer elements to assemble particles in a fluid volume~\cite{RN358}. They found solutions for a selection of 3D traps using the boundary element method. However, the geometric constraints imposed by opposing transducers and inevitable reflections in this closed environment limit the attainable shapes to mode geometries highly dependent on the container shape, i.e. cubic symmetries.

In this article, we present a method to realize the first one-step 3D assembly of matter into arbitrary shapes using ultrasound by combining multiple acoustic holograms. In particular, we show that neither counterpropagating waves nor phase discontinuities are required to trap particles in 3D. Our method thus not only allows assembly into arbitrary 3D forms, but also confers flexibility in the experimental setup as all acoustic sources can be placed on one side of the projected field. That leaves room for experimental intervention, i.e. the addition of sample material, the integration with other processes, and access for optical characterization and imaging. We demonstrate the practicability of our compact 3D fields by assembling different classes of matter, namely solid particles, biological cells and hydrogel beads, into 3D shapes in one shot. 

\section*{Results}
\subsection*{Computation of compact 3D holographic fields}
\begin{figure}[tbp]
    \centering
    \includegraphics[width=1.0\textwidth]{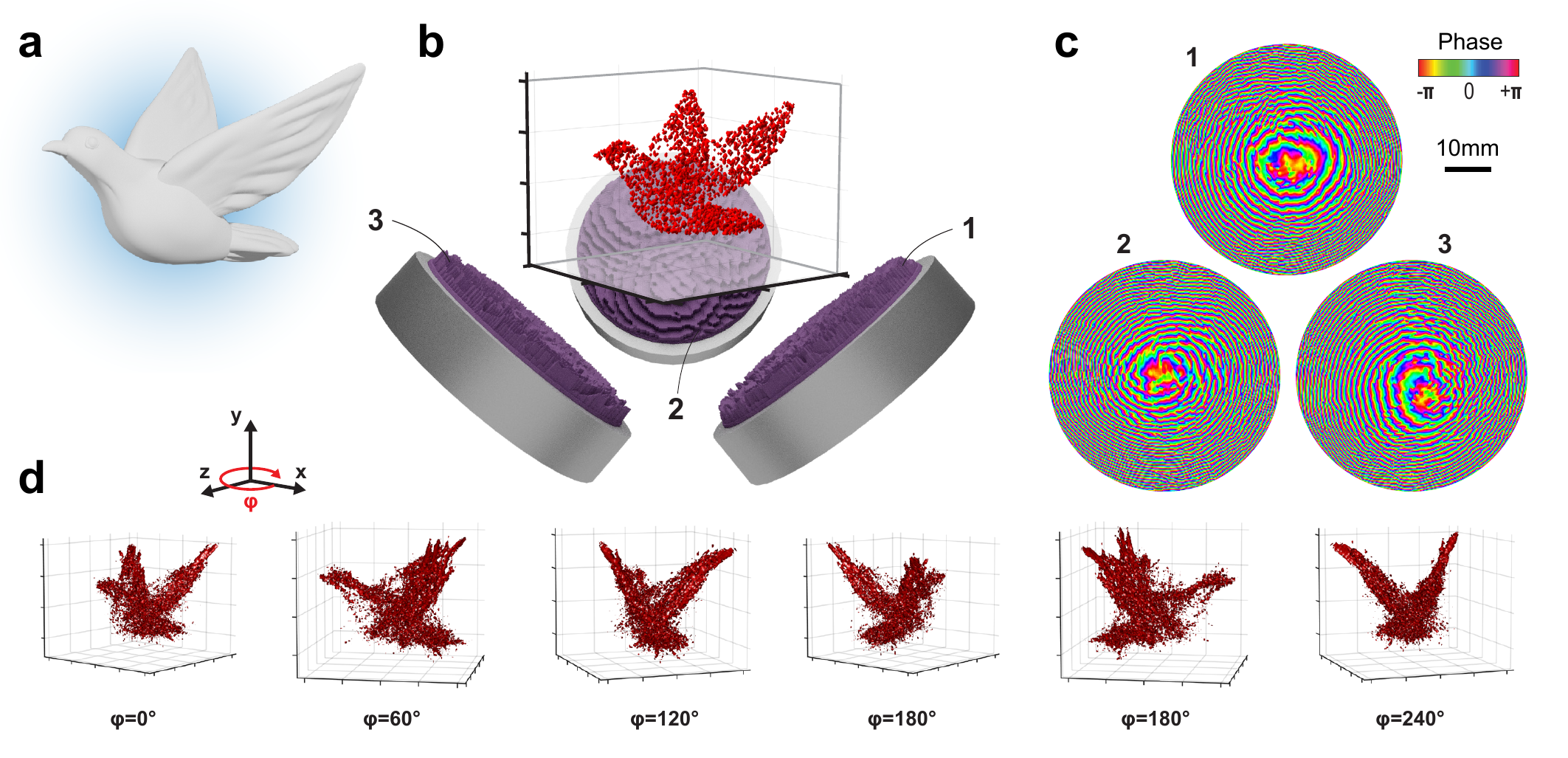}
    \caption{Concept to form compact acoustic 3D pressure images. The target 3D shape \panel{a} is scaled and converted into a voxel representation \panel{b}. The number of necessary holograms as well as their position and orientation relative to the target are chosen as part of the design process. Each hologram is placed on an ultrasound transducer that emits a plane wave at a fixed frequency. \panel{c} Our computer generated holography (CGH) routine finds the optimal phase maps for each hologram. \panel{d} Reconstruction of the ultrasound pressure amplitude emanating from the transducer-hologram sources (1,2 and 3) and projecting into the 3D volume shows good results, displayed here as isosurfaces at 40\% of the maximum pressure.}
    \label{fig:setup}
\end{figure}
The wavelengths of waterborne ultrasound at MHz frequencies are about three orders of magnitude larger than those of visible light. The experimental dimensions and aperture sizes we use in practice, however, are of same order. For this reason the axial and radial resolution scale favorably and we achieve similar resolution in all three spatial directions. Superposition of concurrent fields from different angles leads to high-fidelity acoustic fields due to a wider range of accessible wave vectors (essentially a much larger numerical aperture). Figure~\ref{fig:setup} shows the general concept and workflow. The target object (Fig.~\ref{fig:setup}\panel{a}) is first converted and transferred to the computation volume (Fig.~\ref{fig:setup}\panel{b}). Depending on the problem at hand the number and orientation of the holograms and transducers are chosen and placed relative to the target. In practice, we found orthogonal placement the most versatile. Then we compute phase maps for all holograms using an optimization algorithm (Fig.~\ref{fig:setup}\panel{c}). In this exemplary simulation three transducers with a frequency of \SI{3.5}{\mega\hertz} and a diameter of \SI{50}{\milli\meter} are employed. Results for the generated 3D pressure fields in the shape of the target are shown in Fig.~\ref{fig:setup}\panel{d}. The pressure images in the shape of the 3D target will drive the assembly process.

\begin{figure}[tbp]
    \centering
    \includegraphics[width=0.6\textwidth]{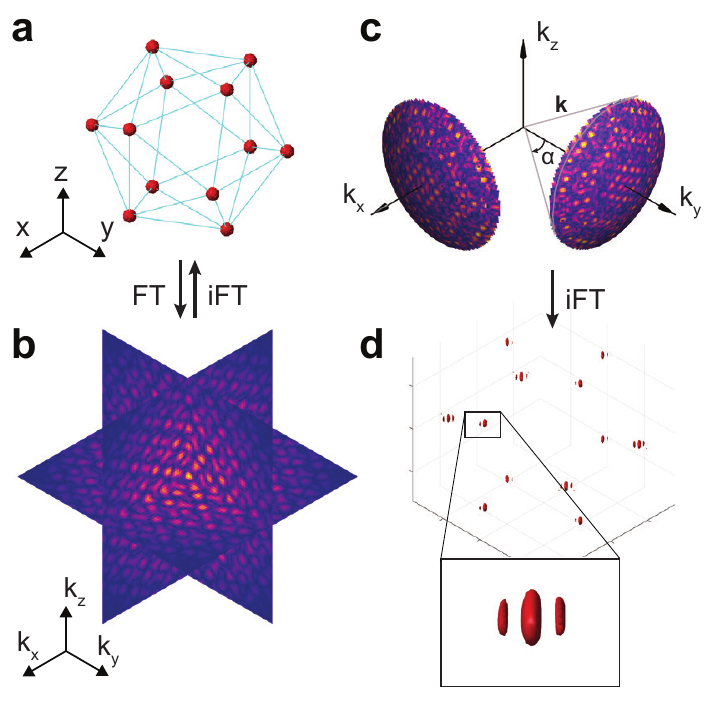}
    \caption{\panel{a} Exemplary 3D arrangement of focal points at the vertex locations of an icosahedron and \panel{b} its spatial frequency spectrum. \panel{a,b} Are linked via the (inverse) 3D Fourier transform, indicated by FT (or iFT). \panel{c} Here, the spectrum is limited to the Ewald surface of a monochromatic wave with wavenumber $k=2\pi/\lambda$ and two sources with limited aperture size (opening angle $\alpha$). \panel{d} Shows the resulting field with close-up view on a single spot.}
    \label{fig:spectra}
\end{figure}
The synthesis of three-dimensional fields forming a target volume or shape is challenging and an active research topic in computer generated holography (CGH)~\cite{RN306,RN608}. The goal of CGH is to find the spatial phase and amplitude distribution of an incident wavefront that will---upon diffraction---reconstruct to form the target image. The goal is achieved by optimizing the output acoustic field under the experimental constraints, such as the incident beam profile and wave shaping modalities. The latter indicates whether the amplitude or the phase (or both) can be controlled. CGH methods for 2D images are well advanced, however accessing the third dimension poses the additional problem that often the optical or acoustic waves have to propagate across regions with highly varying intensities while energy conservation has to be preserved. Some desired 3D target images may thus not be physically realizable, only via a different geometry.
As we show here, an effective way to quickly infer the feasibility of volumetric fields without having to undertake time-consuming volumetric wave propagation calculations is to analyze the 3D spatial frequency spectrum of the target object~\cite{RN585}. Figure~\ref{fig:spectra}\panel{a} shows an exemplary field of focal spots (regions where high intensity is desired) arranged so as to form the vertices of a regular icosahedron. Its 3D Fourier transform (FT) ( Fig.~\ref{fig:spectra}\panel{b}) corresponds essentially to a decomposition into plane waves, each represented by their wave vector \(\vec{k}\). Thus for monochromatic illumination (insonification) with wavelength \(\lambda_0\) this spectrum has to be limited to the surface of a sphere---the Ewald sphere---whose radius is equal to the wavenumber \(|\vec{k}|=k_0=2\pi/\lambda_0\). However, sources of light or sound are restricted in space. The finite aperture of a real source thus further constrains the wave vectors to a cone of opening angle \(\alpha\) which is given by the dimensions of the source (Fig.~\ref{fig:spectra}\panel{c}). Back-transformation of this limited spectrum using the inverse Fourier transform (iFT) reveals one possible solution to the initial field. We call this approach the Fourier Constraint Method (FCM). Alternatively, solutions can be found by varying the complex phases of points in the spatial field, which then determines the hologram that is to be used in conjunction with the source. The same procedure can be extended to multiple transducers with their respective holograms in arbitrary (source) orientations and configurations. An example for the spectrum limited to two holograms along the x and y direction, respectively, can be seen in Fig.~\ref{fig:spectra}\panel{c}. The inverse FT reveals the resulting field in Fig.~\ref{fig:spectra}\panel{d}, where characteristic intensity (pressure) nodes around the foci are visible.

Iterating through these steps to compute a hologram has been termed a 3D variant of the Gerchberg-Saxton (GS) algorithm in optical CGH~\cite{RN586,RN587}. However, this is not directly applicable in the case of ultrasound, which we will use for assembly. In ultrasound, the near-field needs to be considered in contrast to the farfield typically encountered in optics, where the source and target are related by a simple Fourier transform. Before source constraints can be applied it becomes necessary to propagate the wave from the image to the hologram plane. This propagation is accomplished in a straightforward manner using the angular spectrum method applied to the limited spectrum, which is contained on the section of the Ewald surface with regards to the opening angle of the source. The FCM delivers very good results for fields that consist only of collections of focal points. However, one difficulty for it (and including GS and its derived iterative angular spectrum approach, IASA\cite{RN246,RN300}) is to achieve uniform amplitude over extended regions, which is particularly desirable for acoustic traps making up lines, surfaces or volumes. To improve the homogeneity of amplitudes over extended traps we therefore resorted to parallel computation of the volumetric field using the angular spectrum method combined with a gradient-based optimization procedure. Non-convex optimization has been employed previously in this context with pioneering work done by Zhang et al.~\cite{RN616}. In summary, our new approach is suitable for holographic reconstruction in the acoustic near-field and allows for arbitrary numbers of holograms at user-defined positions and orientations to generate complex 3D pressure patterns in space.

\subsection*{Directed 3D assembly of microparticles}
\begin{figure}[tbp]
    \centering
    \includegraphics[width=0.5\textwidth]{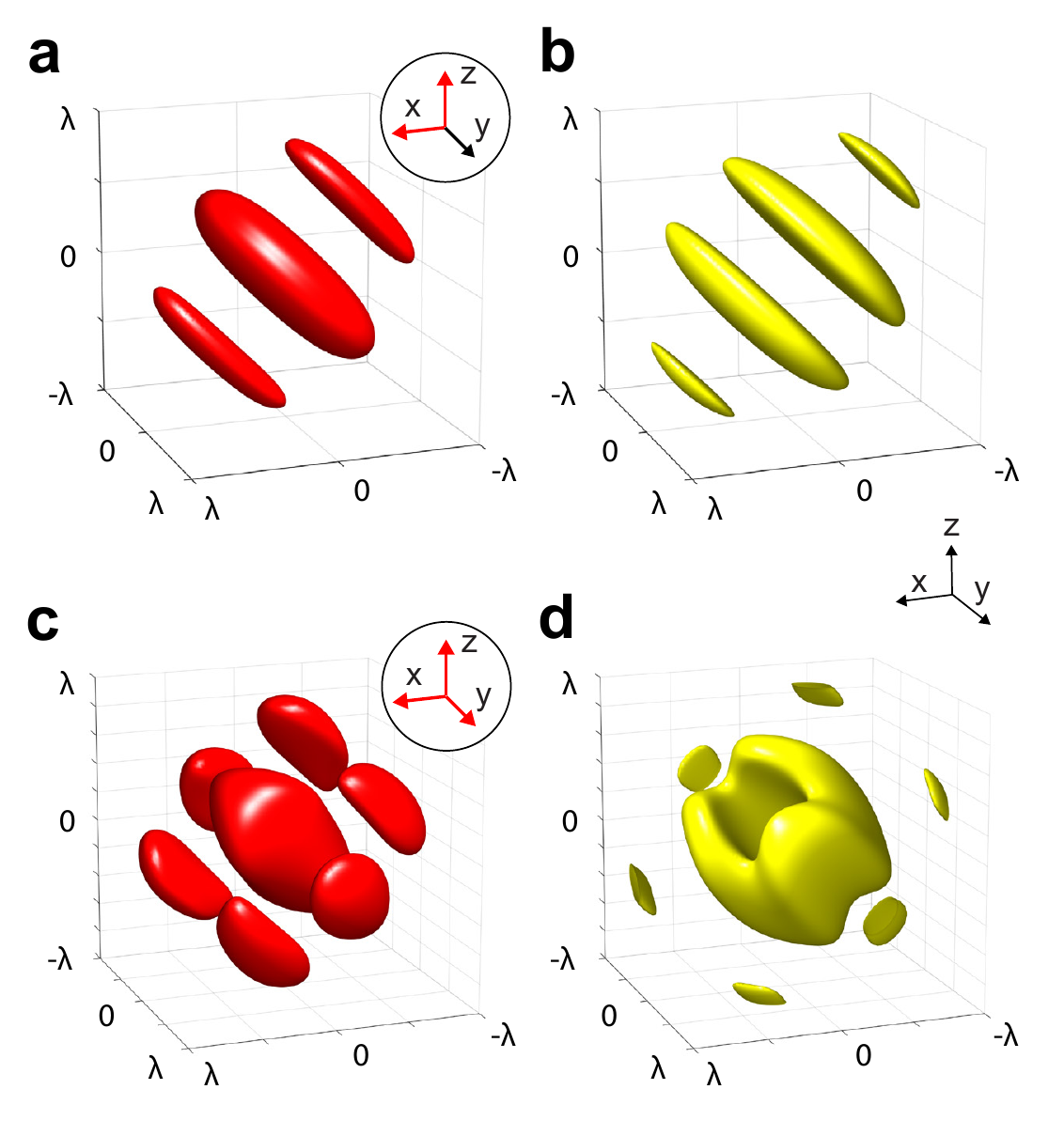}
    \caption{Particle trapping potential around the mutual focal spot of two, respectively, three focused beams. Beam directions are indicated by red arrows in the circular insets. \panel{a} Normalized intensity distribution at the focus of two beams pointed in X and Z direction (isosurfaces at $0.25\cdot\mathrm{max}\{I\}$), and \panel{b} the particle trapping sites indicated by the negative Gor'kov potential $-\Upsilon$ (isosurfaces at $-0.25\cdot\mathrm{min}\{\Upsilon\}$). \panel{c,d} Intensity distribution (isosurfaces at $0.1\cdot\mathrm{max}\{I\}$) and negative Gor'kov potential (isosurfaces at $-0.25\cdot\mathrm{min}\{\Upsilon\}$) for three focused beams in X, Y and Z direction. All dimensions are normalized by wavelength \(\lambda\) and the potentials have been computed for a silica gel particle with radius \(a_\mathrm{p}=\lambda/10\).}
    \label{fig:gorkov-point}
\end{figure}
Structured acoustic fields have long been used to trap and assemble microparticles~\cite{RN316,RN728}. A popular measure to gauge the attainable trapping potential in acoustics is the Gor'kov radiation potential because of its straightforward computation~\cite{RN473}, which has been extended to complex structured fields~\cite{RN241} and arbitrary shaped particles~\cite{RN741}. Figure~\ref{fig:gorkov-point} visualizes two examples of superimposed beams: two (Fig.~\ref{fig:gorkov-point}\panel{a}) or three (Fig.~\ref{fig:gorkov-point}\panel{c}) beams are focused to a point at the origin (0,0,0) coming from orthogonal directions corresponding to the axes highlighted red in the insets. The particles used in this work (e.g. silica gel particles) have positive acoustic contrast against water and are consequently attracted to minima of the Gor'kov potential. Therefore the opposite sign of the potential is displayed as isosurfaces in Figs.~\ref{fig:gorkov-point}\panel{b} and \panel{d}. The closed surfaces visualize the approximate form of particle aggregates in a superimposed focal spot. It should be noted here that these traps do not constitute a single connected potential well that the particles fall into, because the trapping regions span multiple pressure nodes in proximity to the focal points, leading to a striped or clustered appearance of the assembled structures. Using higher frequency ultrasound will reduce the spacing between stripes. % This can be advantageous for fabrication and cell culture as many methods are tolerant to or might even require some spacing between their constituents. Fabrication methods that rely on polymerization for example, or cell culture that incubate and grow after fixation in hydrogel compound.

\begin{figure}[tbp]
    \centering
    \includegraphics[width=0.8\textwidth]{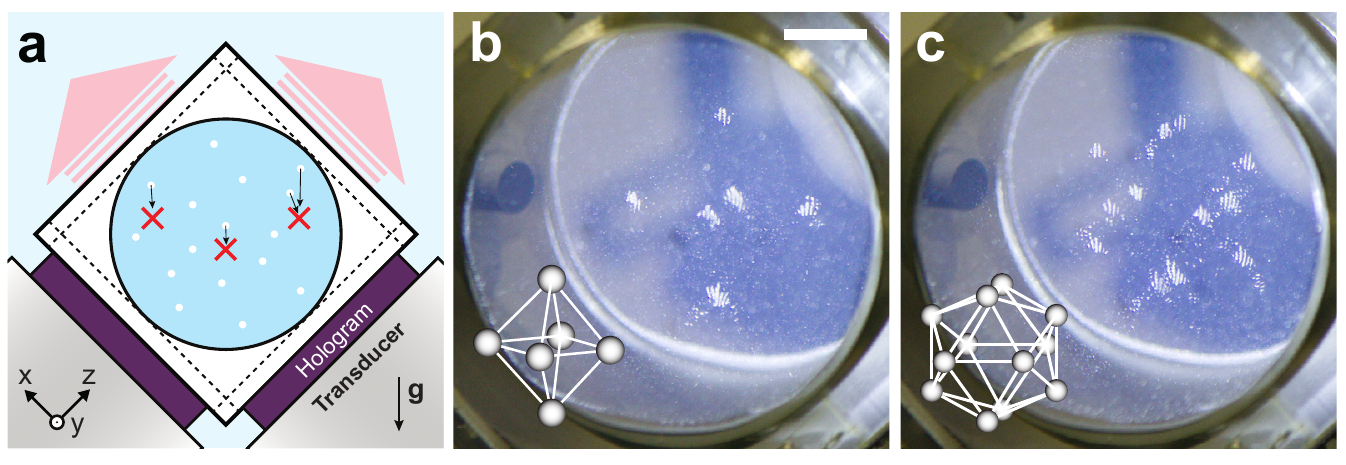}
    \caption{Trapping of silica gel microspheres with \SI{2.25}{\mega\hertz} ultrasound at specified points in 3D. \panel{a} Schematic of the experimental setup using two holograms and a removable particle container. After initial resuspension by shaking the particles (white dots) sediment under gravity $g$ until they pass a trapping site (red cross). The trap locations represent vertices of platonic solids: \panel{b} octahedron, \panel{c} icosahedron. Scale bar \SI{10}{\milli\meter}.}
    \label{fig:platon_assembly}
\end{figure}
Trapping of solid particles in specified locations in 3D has been demonstrated by arranging trapping sites at vertices of the platonic bodies. In this experiment two holograms were computed for two source beams intersecting at \SI{90}{\degree}, as shown in the schematic in Figure~\ref{fig:platon_assembly}\panel{a}. A removable cubic container was filled with a suspension of porous silica gel spheres in water and placed so that it rested directly on the holograms. Its faces were covered with acoustically transparent windows and therefore allowed the ultrasound field to propagate through the container with minimal reflection. The holograms were computed so that an acoustic image forms at the center of the container. At the start of the experiment the cube was shaken manually and then placed on the rig. The microparticles quickly sedimented under the influence of gravity, however, particles that traversed the trapping sites around each focal point were retained by the acoustic radiation force. Figures~\ref{fig:platon_assembly}\panel{b, c} show photographs of the final trapping state for an octahedron and an icosahedron, respectively.

\begin{figure}[tbp]
    \centering
    \includegraphics[width=0.8\textwidth]{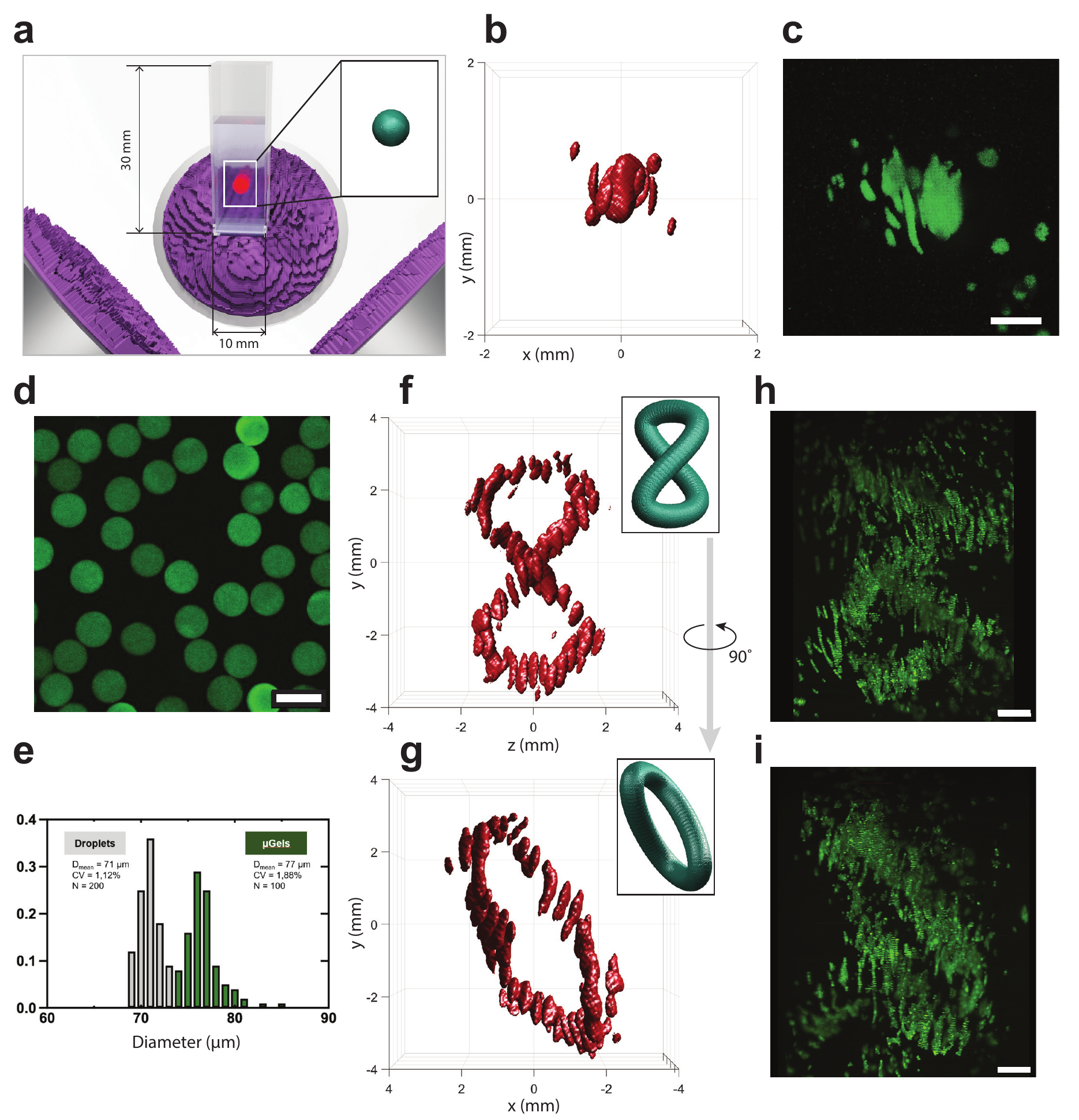}
    \caption{3D holographic assembly of cells and microgel beads. \panel{a} The compact 3D sound image formed by three \SI{2.25}{\mega\hertz} ultrasound transducers and associated holograms is focused into a standard cuvette. The inset shows the target field, in this example an extended spherical volume, where panels \panel{b,c} show the simulated sound intensity I (isosurface at $0.09\cdot\mathrm{max}\{I\}$) and a 3D fluorescence image stack of assembled C2C12 mouse myoblasts, respectively. \panel{d} Microscope image of fluorescent microgels. \panel{e} Size distribution of microgels before and after curing. \panel{f,g} Simulated sound intensity field of a distorted figure eight curve shown in front view and turned \SI{90}{\degree} around the vertical axis. Sound intensity isosurfaces at $0.16\cdot\mathrm{max}\{I\}$. \panel{h,i} Assembly of microgels shown according to views in panels \panel{f,g}. Scale bar of \panel{d} is \SI{100}{\micro\meter}. All other scale bars \SI{1}{\milli\meter}}
    \label{fig:assembly-materials}
\end{figure}
The combination of beams from orthogonal directions (causing highly varying wave vectors) causes interference patterns with pressure nodes that can be used to trap and assemble matter with a positive acoustic contrast. This is useful as most materials, including cells, have a positive acoustic contrast in water (PDMS is an exception). Figure~\ref{fig:assembly-materials} shows assemblies of biological cells (C2C12 mouse myoblasts) as well as microscopic hydrogel beads (gelatin methacrylate, GelMa). The assembled particles and cells were fixed in a slow-curing hydrogel medium, which leaves enough time for excess particles to settle before solidification. The experiments were performed inside square cuvettes, which were aligned to the transducers so that the target field is projected towards the center of the medium (Fig.~\ref{fig:assembly-materials}\panel{a}). The first demonstration is the assembly of biological cells to an extended spherical volume. Panels \panel{b,c} show the simulated sound intensity field and the final cell assembly, respectively. We recorded and confirmed the 3D structure of the sample using a custom-built laser-sheet fluorescence imaging setup.
Hydrogel beads labelled with a fluorescent dye were fabricated on a microfluidic platform. A fluorescence microscope image and their size distribution before (droplets) and after curing (\textmugreek Gels) can be seen in Figs.~\ref{fig:assembly-materials}\panel{d,e}, respectively.
%It can be seen that the hydrogel beads swell by a small amount relative to the droplet size before curing. However, the microfluidic fabrication method leads to a narrow and well-controlled size distribution and regarding future work it further allows to load, e.g., cells into the beads.
We then computed the holograms to obtain a compact figure eight curve and used the same experimental setup to assemble the hydrogel beads. Figs.~\ref{fig:assembly-materials}\panel{f,g} illustrate the simulated sound intensity fields in front view and side view, respectively. Laser-sheet images of the same sample are shown corresponding to the views in Figs.~\ref{fig:assembly-materials}\panel{h,i}.

\section*{Discussion and outlook}
In summary, we introduced a concept for generating compact three-dimensional acoustic pressure shapes through superposition of fields from multiple acoustic holograms. To compute the fields in 3D we initially used a Fourier constraint method (FCM), which is sufficient to generate point-like targets. Further, FCM is useful as a fast and intuitive way to gauge the feasibility whether certain target shapes can be generated with a particular configuration of holograms. In order to obtain traps beyond points that extend along lines in space, one needs to resort to a gradient-based optimization of the phase profile. To this aim, we devised an alternative computational method based on non-convex optimization, that minimizes the mean squared error between the generated field and the target field via a projected gradient descent scheme. Note that closely related approaches to non-convex optimization have been proposed by \textcite{RN616} and subsequently been taken up by \textcite{RN736}. However, these works consider standard single source optical holograms. Here, we show how to obtain complex 3D pressure shapes using multiple acoustic sources and holograms.
The compact acoustic fields we created experimentally using the computed holograms allowed us to instantly assemble matter into arbitrary 3D shapes for the first time. We have demonstrated this concept with silica gel particles, biological cells and also hydrogel beads in setups consisting of two or three transducers, where each is fitted with one hologram. The assemblies were fixed in a hydrogel phase for subsequent analysis. Our compact 3D acoustic fields trap and assemble particles around specified points or along curved lines in a bulk suspension and even inside conventional sample tubes and cuvettes. 

Due to the versatility and ability to accommodate different materials our method shows promise for fabrication of scaffolds or directly assembling biological tissue. We foresee applications of compact high-fidelity three-dimensional ultrasound fields in medical therapy, targeted drug delivery and neuro-stimulation. The effects of ultrasound in these emerging fields have already been shown and are investigated by many research groups~\cite{RN734}. Further, the directed 3D assembly of matter is a promising avenue for rapid prototyping and in particular for tissue engineering.

\section*{Acknowledgements}
%The research was in part supported by the European Research Council under the ERC Advanced Grant Agreement HOLOMAN (no. 788296) and by the Max Planck Society.
The authors thank Dr. Sadaf Pashapour and the microfluidic core facility at the Institute for Molecular Systems Engineering and Advanced Materials (IMSEAM) at Heidelberg University for the microfluidic device fabrication. The research was in part supported by the European Research Council under the ERC Advanced Grant Agreement HOLOMAN (No. 788296); the Max Planck Society and the Fraunhofer Society (AKUSTOGRAMME); as well the German Research Foundation (Deutsche Forschungsgemeinschaft, DFG) under Germany's Excellence Strategy via the Excellence Cluster 3D Matter Made to Order (EXC‐2082/1‐390761711)..

\newpage
\printbibliography

\end{document}

%% file: preamble.tex
\usepackage[utf8]{inputenc}
\usepackage[TS1,T1]{fontenc}

\usepackage{amsmath}

\usepackage{textgreek}

\usepackage[shortcuts]{extdash}

\usepackage{siunitx}
\DeclareSIUnit{\volpercent}{vol\%}
\DeclareSIUnit{\wtpercent}{wt\%}
\DeclareSIUnit{\wtvolpercent}{\%w/v}

\usepackage{graphicx}

\newcommand{\panel}[1]{(\textbf{{#1}})}

\renewcommand\vec{\mathbf}

\usepackage[backend=biber,style=nature]{biblatex}
\usepackage{setspace} 

\newcommand{\ts}[1]{\textsuperscript{#1}}